\documentstyle[12pt,preprint,aps,epsfig,floats,amssymb]{revtex}
\newcommand{\ges}{\mbox{$g_1^2$}}
\newcommand{\gzs}{\mbox{$g_2^2$}}
\newcommand{\gds}{\mbox{$g_3^2$}}

\newcommand{\la}{\mbox{$\lambda$}}
\newcommand{\gt}{\mbox{$g_t$}}
\newcommand{\gts}{\mbox{$g_t^2$}}

\tighten
\begin{document}
\preprint{
\noindent
\begin{minipage}[t]{2in}
\begin{flushleft}
\end{flushleft}
\end{minipage}
\hfill
\begin{minipage}[t]{2in}
\begin{flushright}
IISc-CTS-10/01\\
hep-ph/0104286\\
\vspace*{.2in}
\end{flushright}
\end{minipage}
}
\draft
\title{Higgs Mass in the Standard Model from \\
Coupling Constant Reduction}

\vskip 4cm

\author{B. Ananthanarayan\\
J. Pasupathy}

\vskip 2cm

\address{Centre for Theoretical Studies,\\
Indian Institute of Science,\\
Bangalore 560 012, India.}

\maketitle

\vskip 4cm

\begin{abstract}
	Plausible interrelations between parameters of the
        standard model are studied. The empirical value of the top quark
        mass, when used in the renormalization group equations, suggests
        that the ratio of the colour SU(3) gauge coupling $g_3$,
        and the top
        coupling $g_t$ is independent of the renormalization scale. On
        the other hand, variety of top-condensate models 
        suggest that the Higgs self-coupling $\lambda$ is proportional
        to $g_t^2$.  Invoking the requirement that the 
	ratio $\lambda(t)/g_t^2(t)$ 
	is independent of the renormalization scale
	$t$,  fixes the Higgs
	mass. The pole mass of the Higgs [which differs from the
        renormalization group mass by a few percent] 
        is found to be $\sim 154$ GeV for the one-loop
        equations and $\sim 148$ GeV for the two-loop equations.
\end{abstract}


\newpage

\section{Introduction}
  It is widely believed that not all the parameters occurring in
  the standard model are independent and there must be some interrelation
  between them. The most striking feature of the standard model in this regard
  is the fact that top quark mass is nearly same as the electroweak symmetry
  breaking parameter (EWSB), $m_t\simeq 174$ GeV ($=(2 \sqrt{2}G_F)^{-1/2}$,
  where $G_F$ is the Fermi constant). 
  This has lead several authors to
  propose mechanisms similar to the BCS superconductors for EWSB~\cite{BHL}. 
  It was already 
  noted way back in 1961 by Nambu and Jona-Lasinio~\cite{nambujonalasinio}, 
  that in these models, besides the zero mass
  collective excitations, [which become the longitudinal component of
  the gauge bosons] there is also a scalar bound state of the massive
  quasi-fermions with a mass appproximately twice that of the fermion mass.
  This theoretical prediction was confirmed experimentally in BCS 
  superconductors in Raman spectra studies.
  This generic relation between the fermion mass and the associated scalar
  particles is present in He-III and in spectra of several nuclei where
  pairing place an important role.  It is conceivable, that BCS type 
  symmetry breaking is truly ubiquitous and extends also to electroweak 
  symmetry breaking.  For our purposes it is sufficient to note that
  such theories the scalar self-coupling is proportional
  to the square of the quasi-fermion scalar coupling.

	Bounds on the top-quark mass as well as the Higgs mass 
        were obtained many years ago by Cabibbo et. al.~\cite{cabibbo}
        by requiring that the coupling constants remain positive
        and the vacuum state be stable. This in turn has given rise
        to a large activity which uses phenomenological ideas like
        infrared fixed points, quasi-fixed points and supersymmetric
        extension of these.  For a review of these topics, see e.g., 
        ref.~\cite{schrempp}.

	On the other hand several years ago Zimmermann introduced the 
   idea of reduction of coupling constants in a renormalizable field
   theory~\cite{zimmermann}. 
   It is conceivable that a theory written in terms of several
   coupling constants $\lambda_0 , \lambda_1 ,..., \lambda_n$
    contains really only one
   independent constant $\lambda_0$ with others expressible as functions of
    $\lambda_0,$  
    \begin{equation}
    \lambda_j=\lambda_j(\lambda_0),\,\,\, j=1,...,n.
    \end{equation}
    Zimmermann, then showed that the beta functions must 
    satisfy relations of the type:
\begin{equation}
\beta_0 {d \lambda_j \over d \lambda_0} = \beta_j,
\end{equation}
where $\beta_j$ denoted the $\beta$-function corresponding to $\lambda_j$.

   Assuming a power series expansion, one may try to solve these equations
   and then discover relations between couplings leading to a reduction 
   in the number of independent ones. In the context of the standard model
   this programme has been considered in detail in ref.~\cite{details}. 
   A major difficulty in this programme is the following. If one considers 
   for example the reduction of the 
   U(1) gauge coupling $g_1$ in terms of the
   SU(3) colour coupling $g_3$ 
   in the lowest order, it has no solution,
   which can be traced to the fact that while non-abelian gauge couplings
   have negative beta funnction (asymptotically free), in contrast to
   U(1) which has positive beta function with a coupling that grows
   with energy. 

   The renormalization group equations being first order differential
   equations require for their solution
   the specification of the couplings at some arbitrary
   scale.   In the one-loop approximation,
   the equations for $g_1,\, g_2$ and $g_3$ the gauge couplings,
   and $g_t$, the Yukawa coupling of the
   Higgs to the top quark, can be solved easily using the experimental
   values of the gauge couplings say at the Z-mass and the top quark mass.
   The details are provided in Sec. II.   In this introduction we simply
   note from Fig. 1 where the solution of the coupling constants $g_3^2(t)$
   and $g_t^2(t)$ for an energy scale running from 10 GeV to a TeV is
   displayed, that $g_t^2(t)$ like $g_3^2(t)$
   also falls with increasing energy and the
   ratio of $g_t^2(t)/g_3^2(t)$ roughly remains the same over this range.
   This is discussed in more detail later.

        It was mentioned above that BCS type theories that the Higgs
        self-coupling $\lambda$ is proportional $g_t^2$.   To carry
        this analogy over to a relativistic field theory, we need to
        specify the energy scale where we expect these relations to
        hold.

	Guided by the observation of the near scale independence of
	the ratio of the $g_3^2(t)/g_t^2(t)$  
	made above, we now make a simple
        hypothesis. The ratio $\lambda(t)/g_t^2(t)$ is independent of the 
        renormailzation scale $t$. This fixes the integration constant of
        renormalization group
        equation for $\lambda$ and therefore the Higgs mass.

        In an earlier work, one of us had used this criterion of scale
        independence of the coupling constant ratios to one-loop
        to determine the Higgs mass~\cite{JP}.   
    Here we extend these considerations to two-loop renormalization
    group equations.  In ref.~\cite{JP}, the difference between the pole mass 
    and the renormalization group mass was ignored.  
    Here we repair the deficiency.
    The rest of this paper is organised as follows. In  Sec. II
    we specify our normalisations, the pole mass and renormalization
    group mass relations
    and the solution of renormalization group equations in 
    the one-loop and two-loop cases.
    In Sec. III, we compare our results with
    indirect estimates of the Higgs mass from electroweak precision
    data and also some of the other theoretical ideas on the subject.
    An appendix explains the relation of our assumption of scale
    independence to the coupling constant reduction method.

\section{Renormalization Group Equations and Results}

        In a renormalizable field theory, one first specifies the
        values of the various couplings and masses at some arbitrary
        momentum scale $\mu$. The physical matrix elements are obtained
        from the Green functions. The requirement that the physical
        quantities are independent of $\mu$ leads to the renormalization group
        equations. 

The renormalization group equations up to two-loops in perturbation theory
have been calculated in the $\overline{{\rm MS}}$
renormalization scheme for the couplings of the standard model and
its minimal supersymmetric extension, for a review,
see e.g. ref.~\cite{schrempp}.
In the following we shall consider the solution of the renormalization group 
equations
for the standard model setting all couplings other than
$g_1,\, g_2, g_3, g_t$ and $\lambda$, to zero.  The renormalization
group equations valid for $\mu>m_t$ read
($t=\ln (\mu/m_t)$):
\begin{eqnarray}
  \frac{{\rm d}\,\ges}{{\rm d}\, t}&=&\frac{g_1^4}{8\pi^2}\,
  \bigg(\;\;\;\,\frac{41}{10}\;\;+\frac{1}{16\pi^2}\Big(\frac{199}{50}\ges
  +\frac{27}{10}\gzs+\frac{44}{5}\gds-\frac{17}{10}\gts
  \Big)\bigg)
\label{RG1}\\ \nonumber\\
\frac{{\rm d}\,\gzs}{{\rm d}\,t}&=&\frac{g_2^4}{8\pi^2}\,\bigg(
-\frac{19}{6}\;\;+\frac{1}{16\pi^2}\Big(\frac{9}{10}\ges
+\frac{35}{6}\gzs+12\gds-\frac{3}{2}\gts
\Big)\bigg)
\label{RG2}\\ \nonumber\\
\frac{{\rm d}\,\gds}{{\rm d}\,t} & = & \frac{g_3^4}{8\pi^2}\,
\bigg(-\,\,7\,\;\;+\;\frac{1}{16\pi^2}\Big(\frac{11}{10}\ges
+\frac{9}{2}\gzs-26\gds-2\gts\Big)\bigg)
\label{RG3}\\ \nonumber\\
\frac{{\rm d}\,\gts}{{\rm d}\,t}&=&\frac{\gts}{8\pi^2}\,\bigg(
-\frac{17}{20}\ges-\frac{9}{4}\gzs-8\gds+\frac{9}{2}\gts
\nonumber\\ \nonumber\\ & &
+\frac{1}{16\pi^2}\Big(\frac{1187}{600}g_1^4-\frac{9}{20}\ges\gzs
-\frac{23}{4}g_2^4+\frac{19}{15}\ges\gds+9\gzs\gds-108 g_3^4
+\frac{393}{80}\ges\gts\nonumber\\ \nonumber\\ & &
+\frac{225}{16}
\gzs\gts
+36\gds\gts -12g_t^4 -12\gts\la
+6\la^2\Big)\bigg)
\label{RGt}\\ \nonumber\\
\frac{{\rm d}\,\la}{{\rm d}\,t}&=&\frac{1}{16\pi^2}\,\bigg(
\frac{27}{200}g_1^4+\frac{9}{20}\ges\gzs+\frac{9}{8}g_2^4-\frac{9}{5}
\ges\la-9\gzs\la-6\gt^4
+12\gts\la+24\la^2 \nonumber\\ \nonumber\\ & &
+\frac{1}{16\pi^2}\Big(-\frac{3411}{2000}g_1^6-\frac
{1677}{400}g_1^4\gzs-\frac{289}{80}\ges g_2^4+\frac{305}{16}g_2^6
-\frac{171}{100}g_1^4\gts\nonumber\\ \nonumber
  \\& &+\frac{63}{10}\ges\gzs\gts
  -\frac{9}{4}g_2^4\gts
  +\frac{1887}{200}g_1^4\la+\frac{117}{20}\ges\gzs\la
  -\frac{73}{8}g_2^4\la-\frac{8}{5}\ges\gt^4
  -32\gds\gt^4\nonumber\\ \nonumber\\& &
  +\frac{17}{2}\ges\gts\la
  +\frac{45}{2}\gzs\gts\la
  +80\gds\gts\la+\frac{108}{5}\ges\la^2+108\gzs\la^2
  \nonumber\\ \nonumber\\& &+30\gt^6
  -3\gt^4\la-144\gts\la^2-312\la^3 \Big)\bigg)
\label{RGla}
\end{eqnarray}

Here we solve these coupled differential equations both for the one-loop
as well as the two-loop case.   We have proceeded as follows:
we first take the values of the gauge couplings
$g_1$, $g_2$ and $g_3$  at the Z-peak and evolve them
via the renormalization group equations to the scale $m_t$, with 5 flavors
of quarks.  For scales larger than
$m_t$ we employ the equations with
6 quark flavors.  The values we take at the Z-peak are $g_1=0.462,\,
g_2=0.652,\, g_3=1.221$.  We take the pole mass for the top-quark to be
$174$ GeV.  The relation between between the pole mass $m_t^{\rm pole}$
and the renormalization group mass $m_t(\mu)$ at the scale $\mu$ is
\begin{equation}
m_t(\mu)=m_t^{\rm pole}(1+\delta_t(\mu))
\end{equation}
with $m_t(\mu)=(v/\sqrt{2})\, g_t(\mu)$, $v=246$ GeV.  The
correction term $\delta_t(\mu)$ is described in detail
in ref.~\cite{schrempp}.   When computed numerically we find
it to be small and for the case at hand,
$\delta_t(\mu=m_t)\simeq -0.05$, which in turn corresponds to
a value of $g_t(\mu=m_t)=0.95$.  

With these inputs it is now easy to obtain solutions for $g_1, \, g_2, \,
       g_3$ and $g_t$. It was mentioned in the introduction that both $g_3$
       and $g_t$ decrease asymptotically and their ratio remains a constant.
       This is displayed in Fig. 1, where the functional dependence of
       $g_3^2(t)$ and $g_t^2(t)$ in the interval corresponding to
       $\mu=10$ GeV to 1 TeV is displayed.  
       It is sufficient to note here that in this interval 
       while both $g_3$ and $g_t$ 
       drop by a factor of nearly 2, 
       their ratio approximately remains constant.  
       Later on we also study the behaviour of this ratio for higher energy 
       intervals in the one-loop and two-loop cases.  It will be
       seen that the constancy improves at the two-loop level as compared
       to the one-loop level.

Now, to determine $\lambda$ we proceed as follows. 
      Consider an arbitrary value at the scale $\mu=m_t$
      for  $\lambda(0)$.        Given this 
      the corresponding value of $\lambda(t)$ is fixed from the 
      renormalization group equation for
      $\lambda(t)$, in the entire domain of $t$ where the equations
      are valid. 

We solve these equations numerically and consider the ratio
$R(t)=\lambda(t)/g_t^2(t)$ conveniently normalized to unity at $t=0$:
$R(t)/R(0)$.  
The results of the computations
are plotted for values of $\mu$ between $m_t$ and a large scale, say
$10^{10}$ GeV, which then
corresponds to $t: 0<t < 17.87$ in Fig. 2.  
We find that the value of
$\lambda(0)=0.176$ corresponding to the middle line is the one
that best satisfies our criterion of constancy of $R(t)$.  
The behaviour of $R(t)$ when $\lambda(0)$ is varied from
slightly from $0.176$ is also shown in Fig. 2.  
Clearly even such small departures of $\lambda(0)$ from $0.176$
lead to significant $t$ dependence and the corresponding ratio of couplings
fail to meet our criterion of scale independence.  
Larger variations would lead to even
larger departure from the desired constancy.
It should be stressed 
here that the best value for $\lambda(0)$ determined by the scale
independence criterion does not depend significantly on the range of $t$.

Starting from this best value $\lambda(0)=0.176$, the corresponding
$\lambda(t)$ is used to determine the Higgs mass as follows.  First
the renormalization group mass is found from the equation
\begin{equation}
m_H(\mu)=\sqrt{2 \lambda(\mu)} v.
\end{equation}
The pole mass is related to the renormalization group equation
mass above by 
\begin{equation}
m_H(\mu)=m_H^{\rm pole}(1+\delta_H(\mu)).
\end{equation}
Details of $\delta_H(\mu)$ are given in the review,
ref.~\cite{schrempp}.  Numerically again it is small and we find it to be
$\simeq-0.05$, for $\mu=m_H$.  Using this, we find that at one-loop 
\begin{equation}
m_H^{\rm pole}= 154 {\rm GeV} \, \, {\rm (one-loop)}.
\end{equation}
Comparing this value of 154 GeV to the value given in ref.~\cite{JP} of
160 GeV, the small difference arises due to the following improvements.
Here we have taken into account the pole mass of the top-quark to be
174 GeV unlike the earlier work where it was the renormalization group mass.
In the earlier work the $t=0$ corresponded to $\mu=m_Z$ and
the fact that only five flavours are operative below $m_t$ was ignored.
Furthermore, the value for the Higgs mass quoted here corresponds to 
its pole mass, which differs by about 5\% from its renormalization group
mass.

In Fig. 3, the analogous computation is performed with the 
two-loop beta functions.   The best value of $\lambda(0)$ changes slightly
to $0.162$.  Repeating the calculations described above for the pole mass,
we find at two-loop
\begin{equation}
m_H^{\rm pole}= 148 {\rm GeV}\, \, {\rm (two-loops)}.
\end{equation}

We now return to the ratio of $R(t)=g_3^2(t)/g_t^2(t)$ 
which we observed in the introduction to be nearly constant.
We plot this ratio, again normalized to its value at $t=0$ for
the range of values for $\mu$ from $m_t$ to $10^{10}$ GeV, in Fig. 4.
It is seen that at one-loop level that this is nearly constant,
with the ratio tending to increase slightly for increasing energies.
In Fig. 5 we show the results for the same
ratio at the two-loop level.  It is observed that the ratio
is unity to within as little as 5\% even at energies as high
as $10^{10}$ GeV.  Note that this feature is virtually independent
of the Higgs mass when varied around our preferred
value of 148 GeV by a few percent on either side and the
figures corresponding to these variations in the Higgs mass lie
on top of each other (see caption of Fig. 5).

      This leads to the question whether one can ``determine'' the top mass 
      by the requirement of scale independence of the ratio 
      $g_t^2(t)/g_3^2(t)$.
      In order to address this, instead of considering a variety of values
      for the top-quark pole mass and the corresponding changes in $g_t(\mu)$,
      we can simply illustrate this equivalently by varying the 
      values of $g_t$ 
      at $\mu =174$ GeV.  We have chosen the
      values 0.85, 0.95 and 1.05.  
      It is seen from Fig. 6 that at one-loop level 
      the scale independence criterion would put the top-quark mass
      at a slightly lower value than the experimental one.
      On the other hand, from Fig. 7 we see that
      at the two-loop level this criterion
      leads to a value for the top-quark mass quite close to the
      experimental value, when $m_H$ is in the neighbourhood of
      148 GeV.  
      
      It is thus seen that our criterion of scale independence
      of the ratios of couplings has an internal
      consistency.  At the one-loop
      level $g_3(t)$ and $g_t(t)$ are independent of $\lambda$
      and the ratio $g_3^2(t)/g_t^2(t)$ is not quite constant.
      However, remarkably
      at the two-loop level both the ratios $\lambda(t)/g_t^2(t)$
      and $g_3^2(t)/g_t^2(t)$ maintain scale independence over a
      significant range of energies.   
      
\section{Discussion}

  Turning now to experiments, one has bounds on the Higgs mass from precision
  electroweak data on the W-mass, $\sin^2\theta_W$, and leptonic width of Z,
  as recently reviewed by Marciano~\cite{marciano}:
  \begin{eqnarray}
  &   \displaystyle m_H=53^{+77}_{-40}\, \, \,{\rm GeV \, (from \,} m_W) & 
  \nonumber \\
  &   \displaystyle m_H=67^{+45}_{-27}\,\, \, {\rm GeV \, (from \, \sin^2
  \theta_W)} & 
  \nonumber \\
  &   \displaystyle m_H=208^{+340}_{-180}\,\, \, {\rm GeV \, (from }\, 
  Z\rightarrow l^+ l^-(\gamma)) & 
  \nonumber 
  \end{eqnarray}
One also has the LEPII bound
\begin{eqnarray*}
m_H > 106 \, {\rm GeV}
\end{eqnarray*}

  On the other hand a Baysean analysis combining the data from
  direct and indirect searches for the Higgs 
  has been done by D'Agostini and Degrassi~\cite{dagostini}, 
  which leads them to expect Higgs mass to be around 
  160-170 GeV with an uncertainty of about 50 to 60 GeV.

  In the present work we have not tried to incorporate, the standard
  model in a more encompassing theory, several of which have been 
  proposed over the last three decades.
  
  Composite models (for a review, see e.g. ref.~\cite{chivukula})
  tend to suggest a larger Higgs mass than the
  values obtained in the present work.   While it is true that
  the Higgs mechanism for explaining spontaneous symmetry breaking
  was introduced in analogy with condensed matter physics, it is
  conceivable that there may be other settings.  There are many
  geometrical ideas in the subject which have been proposed in the
  last three decades.  Here we shall limit ourselves to brief comments
  about some results which are phenomenologically interesting.

  For example in a version of noncommutative geometry,
  considered by Okumura~\cite{okumura}  the Higgs field along
  with the gauge field appears as a connection on $M_4\times Z_N$,
  where $M_4$ is the usual spacetime manifold and $Z_N$ is a discrete
  space.  Okumura obtains the coupling constant relation at the
  unification scale
  \begin{eqnarray}
  & \displaystyle g_2^2=g_1^2=4 \lambda\neq g_3^2 .& \label{oku}
  \end{eqnarray}
  In view of the last inequality there is no grand unification in this
  model.  Extrapolating eq.~(\ref{oku}) to lower energies, yields the
  result  
\begin{eqnarray*}
 m_H\simeq 158\, {\rm GeV} .  
\end{eqnarray*}
 for the renormalization group mass.

	In another class of theories considered for example by 
	Fairlie~\cite{Fairlie} and by Manton~\cite{Manton}, the
	Higgs field makes its appearance from dimensional reduction.
	A gauge field which should transform as an anti-symmetric tensor
	under $SO(N-1,1),\, (N>4)$ when reduced with respect to the
	Lorentz group $SO(3,1)$ can contain Lorentz scalars just as
	a Lorentz four vector when reduced with respect to rotations
	contains a three vector and a three scalar.  Reducing for
	example a gauge field theory in 6 dimensions
	with internal symmetry of the graded Lie group
	$SU(2|1)$ Fairlie found $\sin^2\theta_W=1/4$ and a Higgs mass
	of 426 GeV.  In these models, the Higgs self-coupling arises
	in a manner similar to the quartic coupling of the gauge
	field from invariance requirements in higher dimensions.

  In view of the fact that superstring theories imply higher dimensional 
  theories, these ideas in which Higgs makes its presence due to reduction
  of gauge connections in higher dimensions may be worth re-examining.
  
  Yet another class of interesting theories is due to 
  Roepstroff and Vehns~\cite{roepstorff}.
  In these theories, the Higgs plays the mathematical role of a 
  super-connection which is an extension of the idea of a
  usual connection for a gauge field on space-time manifolds.
  Here again, relations between gauge couplings $g_1,\, g_2$ and
  $\lambda$ are obtained.  For example, in the absence of fermions
  the model predicts $m_H=2 m_Z$, whereas a refined version which
  includes fermions as well yields a value of $m_H \approx 160$ GeV.

	Of course, currently most of the phenomenological activity is
  centered around the minimal supersymmetric extension of the standard
  model.  While tree level relation puts the lightest Higgs to be below
  the Z-mass, radiative corrections puts it at a higher value. 
  If one solves the standard model renormalization group 
  equations at two-loops as we have done here, and identifies the Higgs
  of the standard model with the lightest Higgs of the minimal supersymmetric
  standard model,  
  one may wish to explore plausible values of the
  supersymmetry scale $M_{SUSY}$ as follows.  Given $m_H$, $\lambda$ is
  determined as a function of $t$ and one may want to find out at what
  value of $t=\ln(M_{SUSY}/m_t)$ the tree-level relation: 
  \begin{eqnarray*}
\lambda(M_{SUSY})={1\over 8}({3\over 5} g_1^2(M_{SUSY}) + g_2^2(M_{SUSY}))
\cos^2 2 \beta,
  \end{eqnarray*}
  is satisfied.   If the Higgs mass
  is varied from 115-155 GeV 
  we find (setting $\cos^2 2 \beta=1$), that $M_{SUSY}$ varies from something
  of order a TeV at the lower mass end, to a few thousand TeV at the
  higher end.  It would be interesting to carry out a detailed analysis
  including the effects of the full supersymmetric spectrum.

In summary, it is interesting to note that the scale independence
hypothesis  leads us to expect a Higgs mass of about 150 GeV, and this
value is quite consistent with the limits obtained from precision
electro-weak data.   Experimental discovery of the Higgs and the
determination of its mass at the Tevatron in this range remains
a distinct possibility.  Should the Higgs turn out to have a mass
of around 150 GeV,  it would strongly support the scale independence
of the coupling constant ratios.

\bigskip

\noindent{\bf Acknowledgements:}  JP thanks J. D. Bjorken, T. Ferbel
and V. Srinivasan for their encouragement and valuable correspondence.

\newpage

\noindent{\bf Appendix}

\bigskip

The following is a brief summary of the work of W.~Zimmermann and 
collaborators.  Details can be found in ref.~\cite{zimmermann,details}.

Consider a renormalizable field theory involving coupling constants
$\lambda_0, \lambda_1,\ldots \lambda_n$.  A functional relation between
coupling constants can arise when there is a symmetry in the Lagrangian or 
even otherwise.  Assume now that all couplings are
asymptotically free, i.e., $\lambda_0,\lambda_1,\ldots, 
\lambda_n$ tend to zero
as the renormalization scale tends to infinity, 
and fundamental relations  of the type
      \begin{eqnarray}
&   \displaystyle   \lambda_i = \lambda_i(\lambda_0) \, \, i= 1,2,\ldots n \ & 
\displaystyle ~~~~~~~~~~~~~~~~~~~~~~~~~~~~~~(A.1) \nonumber
      \end{eqnarray}
make it possible to express the $\lambda_i(i=1,\ldots n)$ in terms of 
$\lambda_0$.  Consider any arbitrary Green function of the theory.  
One may study the renormalization group transformation property of 
the Green function first in terms of the original theory where 
$\lambda_0, \lambda_1,
\ldots \lambda_n$ are all regarded as 
independent and the reduced theory written 
in terms of $\lambda_0$.   Comparing the two, consistency demands
the relation
     \begin{eqnarray}
 & \displaystyle    \beta_{\lambda_{0}}\;
     \frac{d\lambda_i}{d\lambda_0} = \beta_{\lambda_{i}} & \displaystyle
 ~~~~~~~~~~~~~~~~~~~~~~~~~~~~~~~~~~~~~~~~~~~(A.2) \nonumber
     \end{eqnarray}
between the beta functions of the theory.

A number of interesting relations following from eqns.~$(A.1)$ and $(A.2)$ 
have been studied in ref.~\cite{details}.  
In the context of the standard model however one 
runs into difficulty, since  $g^2_1$ the coupling
corresponding to $U(1)$ groups is asymptotically increasing.  
In fact, consider a reduction program of $g^2_1$ in
terms of $g^2_3$ in a power series
     \begin{eqnarray}
     & \displaystyle
     g^2_1 = C_0 g^2_3 + C_1 g^4_3+ \;\ldots & \displaystyle 
     ~~~~~~~~~~~~~~~~~~~~~~~~~~~~~~~(A.3)
     \nonumber
     \end{eqnarray}
For three generations the renormalization group 
equations for $g^2_1$ and $g^2_3$ read in the
one-loop approximation as
     \begin{eqnarray}
 & \displaystyle
    \frac{dg^2_1}{dt} =\frac{g^4_1}{8\pi^2}(\frac{41}{40}), & 
     \displaystyle ~~~~~~~~~~~~~~~~~~~~~~~~~~~~~~~~~~~~~~~~~(A.4) \nonumber \\
  & \displaystyle   
  \frac{dg^2_3}{dt} =\frac{g^2_3}{8\pi^2}(-7).  & \displaystyle
  ~~~~~~~~~~~~~~~~~~~~~~~~~~~~~~~~~~~~~~~~~(A.5)
  \nonumber
      \end{eqnarray}
Using $(A.3),\, (A.4)$ and $(A.5)$ in $(A.2)$ finds for the coefficient $C_0$
     \begin{eqnarray}
   & \displaystyle  C^2_0\;\frac{41}{10} = -7\;C_0 & \displaystyle
   ~~~~~~~~~~~~~~~~~~~~~~~~~~~~~~~~~~~~~~~~~(A.6) \nonumber
   \end{eqnarray}
which has solutions (a) $C_0=0$; trivial, or
(b) $C_0=-70/41$; negative and therefore unacceptable.

This difficulty is circumvented in ref.~\cite{details}
by discarding the reductions hypothesis 
for $g_1$ and in view of its smallness in relation to $g_3$
the top coupling $g_t$ and the Higgs self-coupling  
$\lambda$ incorporate it as a perturbation on other reduction equations.
These procedures lead to very low values for the top-quark mass.
Details can be found for example in ref.~\cite{details}.

It is interesting to compare the reduction hypothesis, eq.~$(A.2)$ 
in relation to our approach in this paper where we have
required the criterion:
     \begin{eqnarray*}
& \displaystyle
     \frac{d}{dt}\left(\lambda(t)/g^2_t(t)\right) = 0,& \displaystyle
     ~~~~~~~~~~~~~~~~~~~~~~~~~~~~~~~~~~~~~(A.7)
     \end{eqnarray*}
it follows from eq.~$(A.7)$
     \begin{eqnarray*}
& \displaystyle
     \beta_{\lambda}\;\frac{g^2_t(t)}{\lambda(t)} = \beta_{g_{t}}.
     & \displaystyle ~~~~~~~~~~~~~~~~~~~~~~~~~~~~~~~~~~~~~~~~~~(A.8)
     \end{eqnarray*}
Comparing eq.~$(A.8)$ with eq.~$(A.2)$ we see that our criterion
of scale independence, $(A.7)$ is consistent with the reduction hypothesis if
     \begin{eqnarray*}
& \displaystyle
      \frac{dg^2_t}{d\lambda} = \frac{g^2_t}{\lambda}.&
      \displaystyle ~~~~~~~~~~~~~~~~~~~~~~~~~~~~~~~~~~~~~~~~~~~~~~(A.9)
      \end{eqnarray*}
Clearly if $(A.7)$ is valid, then $(A.9)$ is valid as well.

Thus our scale independence condition is consistent with the 
reduction hypothesis but is not demanded by it.
On the other hand, we have seen in Fig. 7 that the ratio of
the strong coupling $g_3$ and the coupling $g_t$ is scale independent
and in fact could be used to determine
the top-quark mass within its experimental limits.  It is therefore
conceivable that the Higgs coupling is also following a similar pattern.

\newpage

\begin{center}
\vskip 2cm


\begin{figure}
\epsfig{figure=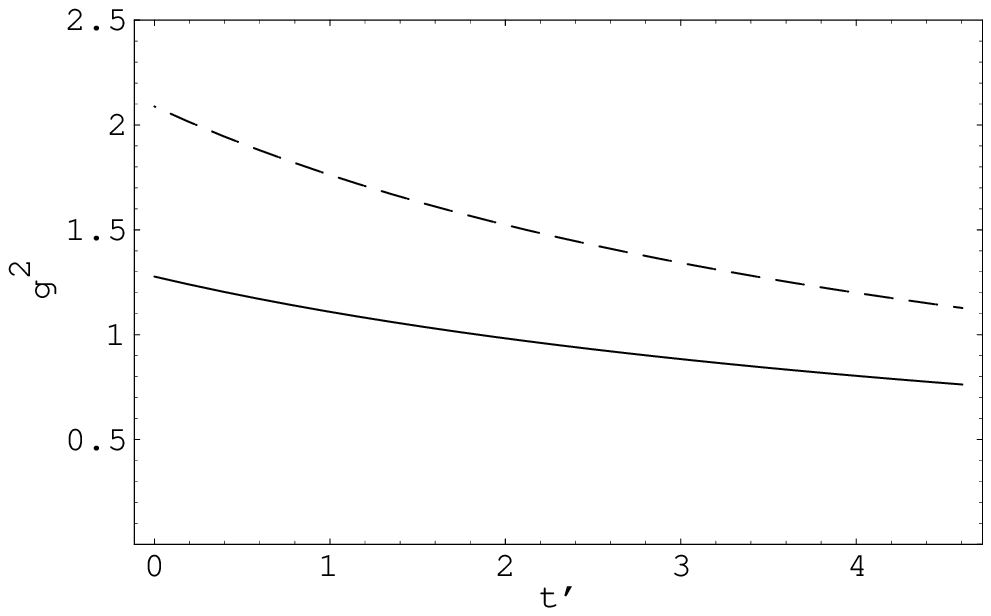,width=15cm,height=15cm}
\caption{Plot of $g_t^2$ and $g_3^2$, 
$t'=\ln(\mu/10)$ at one-loop (solid and dashed respectively),
corresponding to $\alpha_S(M_Z)=0.119,\, m_t^{\rm pole}=174$ GeV,
see Sec. II for details.}
\end{figure}


\begin{figure}
\epsfig{figure=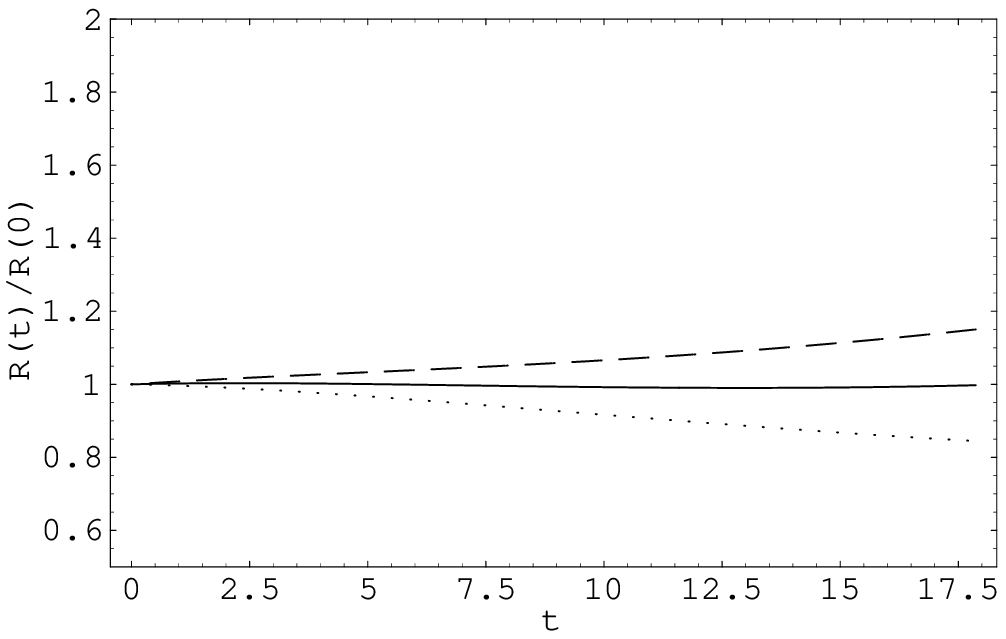,width=15cm,height=15cm}
\caption{Plot of $R(t)/R(0)$ vs. $t$, $R(t)=\lambda(t)/g_t^2(t)$,
$t=\ln(\mu/m_t)$ at one-loop, 
with $\lambda(0)=0.171,\, 0.176,\, 0.181,$
$g_t(0)=0.95$ (dotted, solid and dashed respectively).}
\end{figure}
\vskip 2cm


\begin{figure}
\epsfig{figure=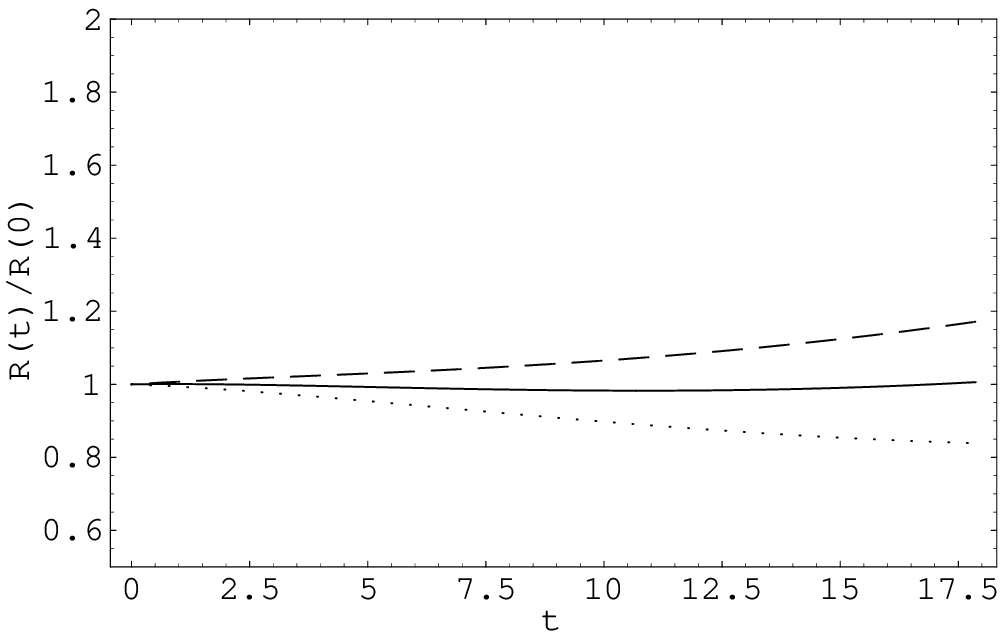,width=15cm,height=15cm}
\caption{Plot of $R(t)/R(0)$ vs. $t$, $R(t)=\lambda(t)/g_t^2(t)$,
$t=\ln(\mu/m_t)$ at two-loops,
with $\lambda(0)=0.157,\, 0.162,\, 0.167$,
$g_t(0)=0.95$ (dotted, solid and dashed respectively).}
\end{figure}
\vskip 2cm


\begin{figure}
\epsfig{figure=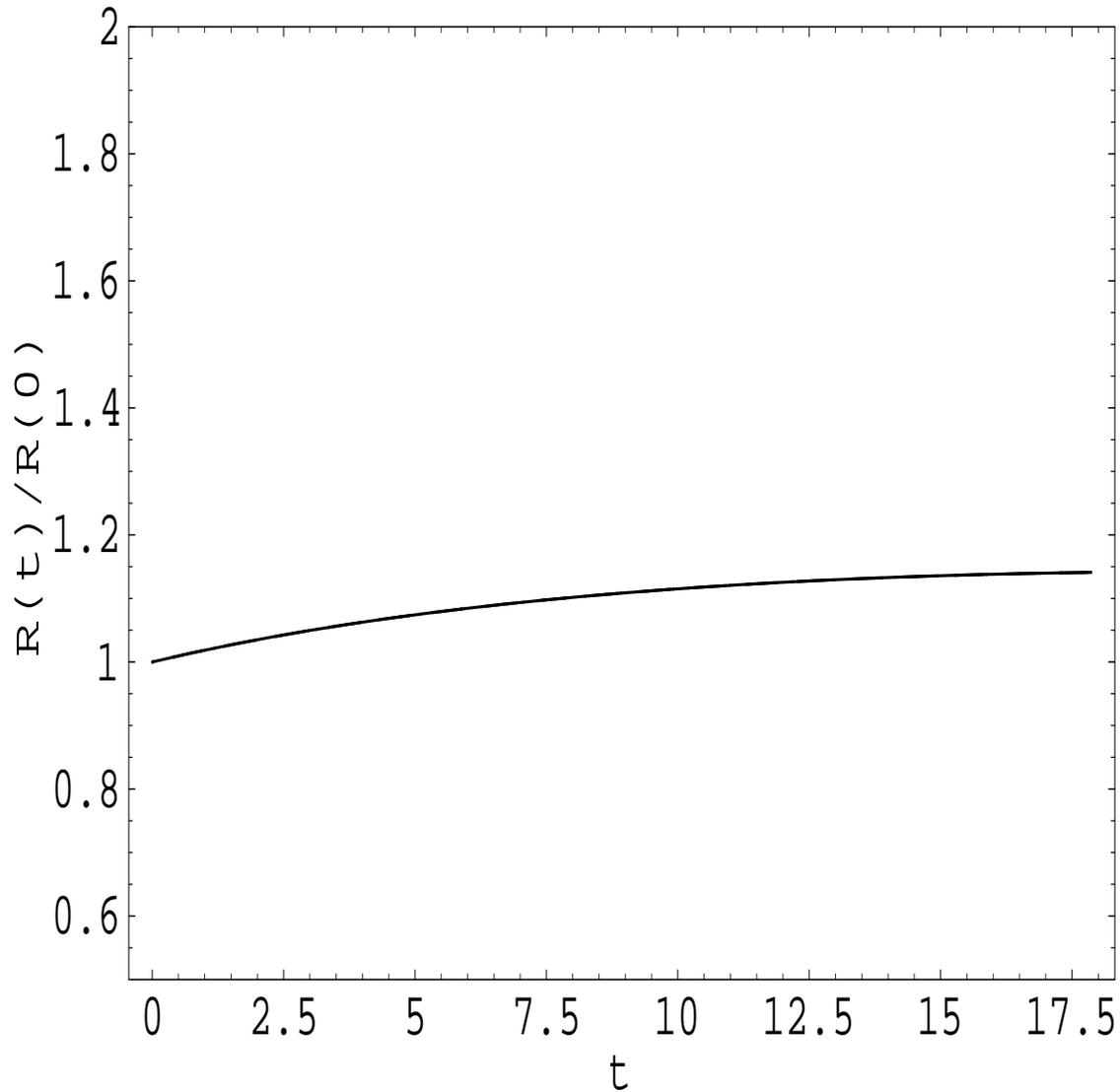,width=15cm,height=15cm}
\caption{Plot of $R(t)/R(0)$ vs. $t$, $R(t)=g_3^2(t)/g_t^2(t)$,
$t=\ln(\mu/m_t)$ at one-loop, 
with $g_t(0)=0.95$ (at one-loop the equations are independent of
$\lambda$).}
\end{figure}


\vskip 2cm
\begin{figure}
\epsfig{figure=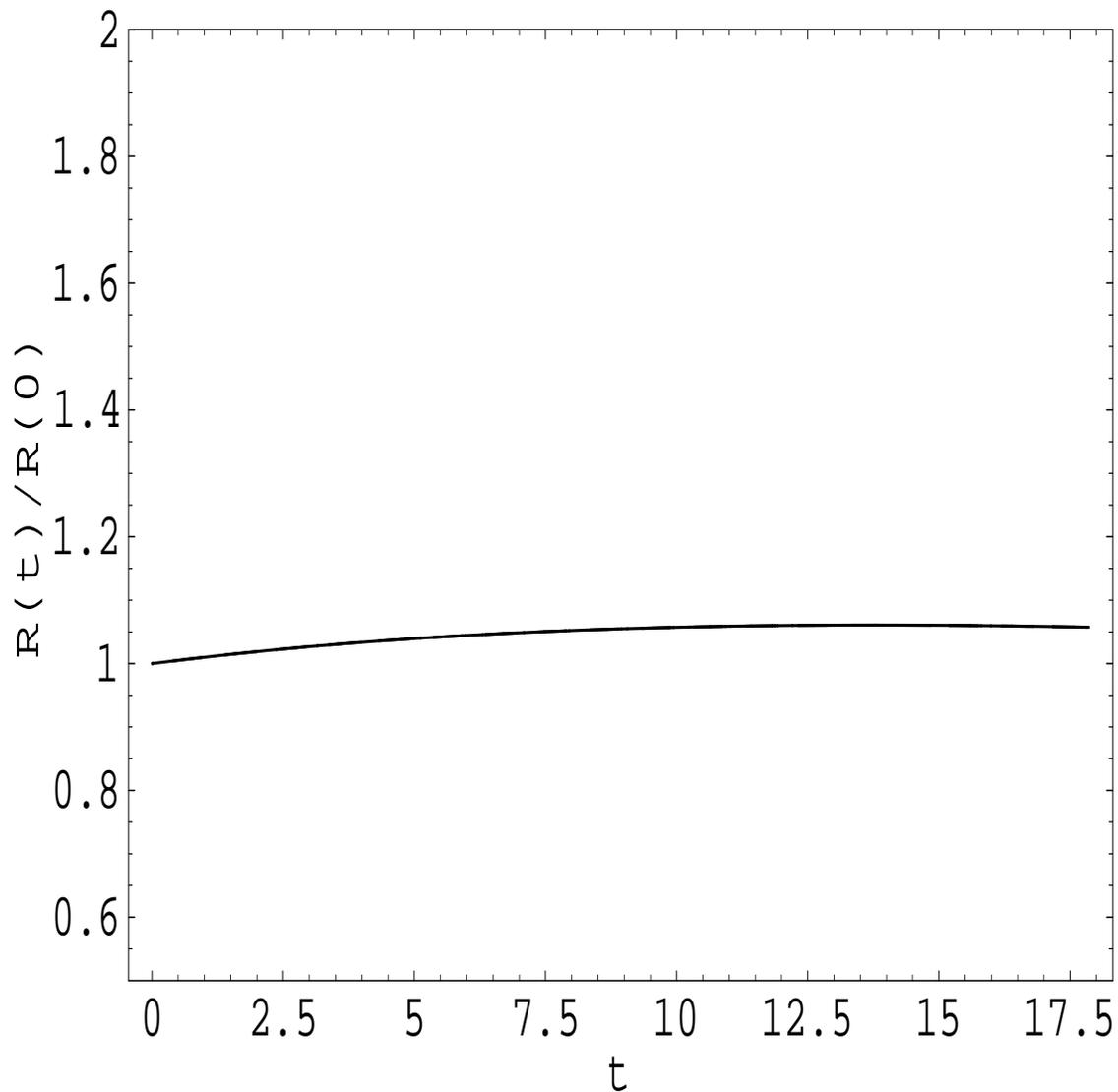,width=15cm,height=15cm}
\caption{Plot of $R(t)/R(0)$ vs. $t$, $R(t)=g_3^2(t)/g_t^2(t)$,
$t=\ln(\mu/m_t)$ at two-loops,
with $g_t(0)=0.95$ and $\lambda(0)=0.162$.  Varying $\lambda(0)$
to $0.157$ or $0.167$ has insignificant effect on this ratio,
and the corresponding curves practically lie on top of the curve
in the figure.}
\end{figure}
\end{center}
\vskip 2cm
\begin{center}
\begin{figure}
\epsfig{figure=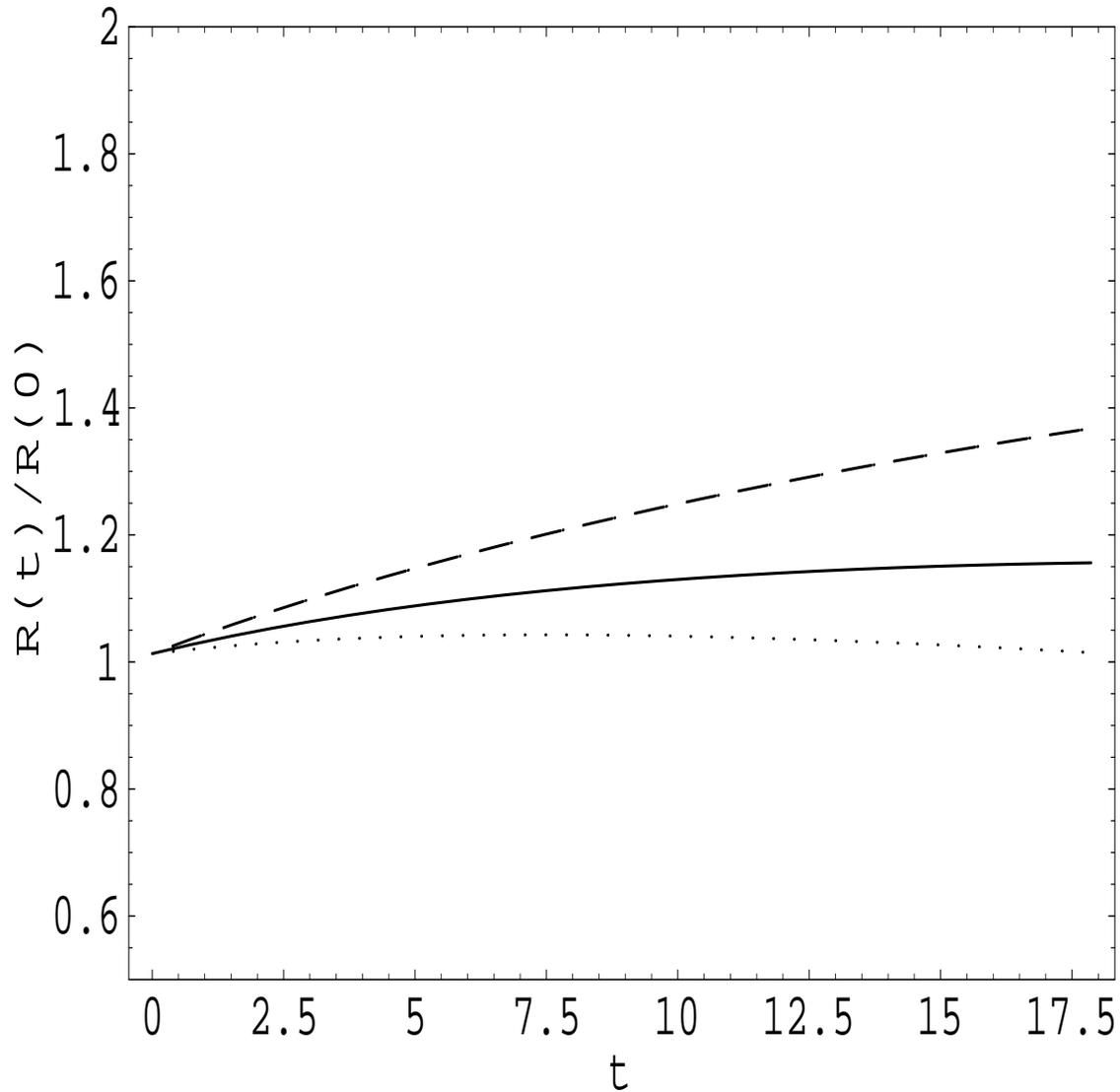,width=15cm,height=15cm}
\caption{Plot of $R(t)/R(0)$ vs. $t$, $R(t)=g_3^2(t)/g_t^2(t)$,
$t=\ln(\mu/m_t)$ at one-loop, with
$g_t(0)=0.85,\, 0.95,\, 1.05$ (dotted, solid and dashed respectively).}
\end{figure}
\vskip 2cm


\begin{figure}
\epsfig{figure=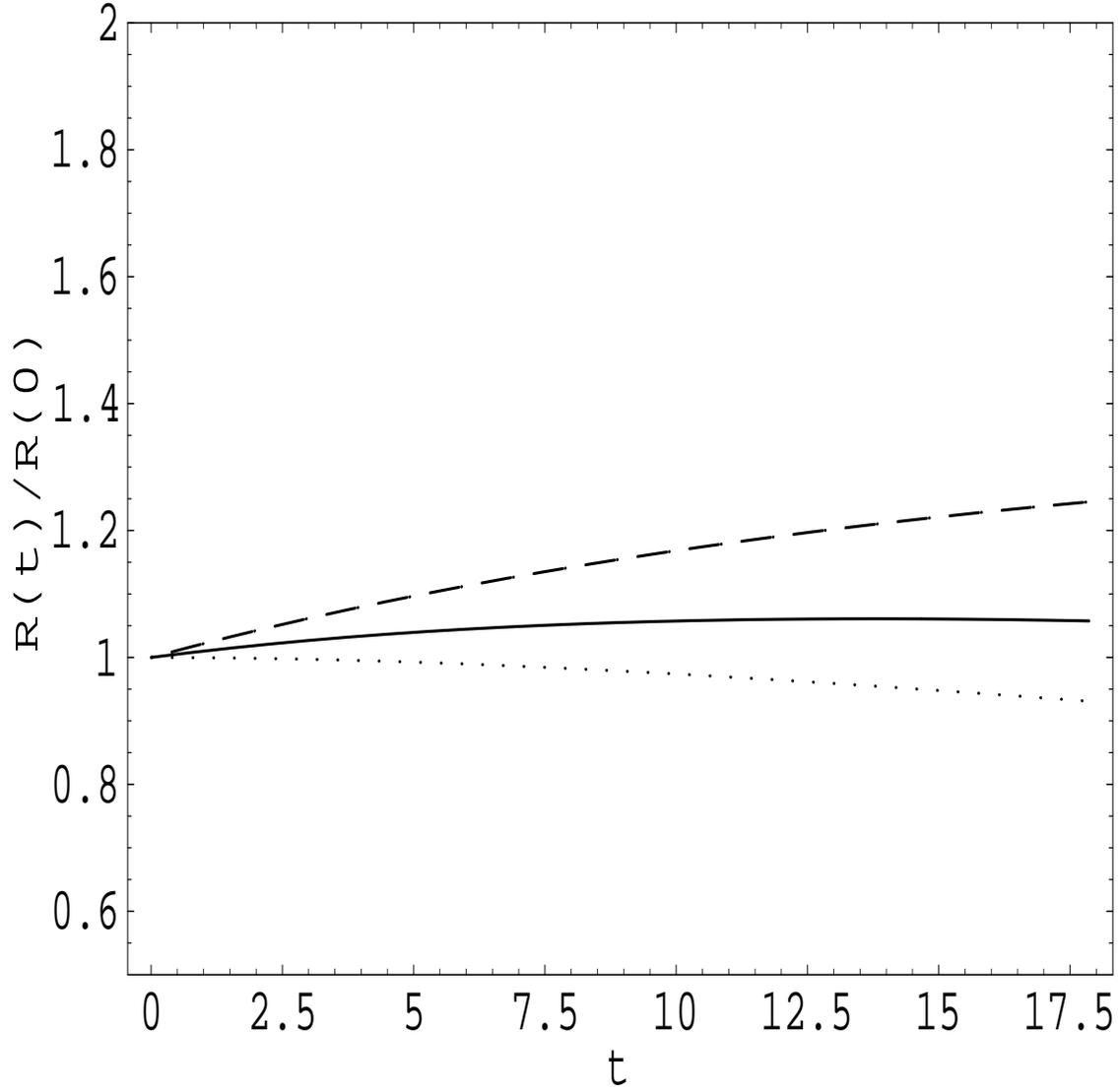,width=15cm,height=15cm}
\caption{Plot of $R(t)/R(0)$ vs. $t$, $R(t)=g_3^2(t)/g_t^2(t)$,
$t=\ln(\mu/m_t)$ at two-loops, 
with $\lambda(0)=0.162$,
$g_t(0)=0.85,\, 0.95,\, 1.05$ (dotted, solid and dashed respectively).}
\end{figure}
\end{center}
\end{document}